# Morphology-, Noise-, and Resolution-Robust Ultrasound Elasticity Imaging with Fourier Neural Operators

*Heekyu Kim*[1], *Hugon Lee*[1], *Minwoo Park*[1], *and Seunghwa Ryu*[1,2*]

[1] Department of Mechanical Engineering, Korea Advanced Institute of Science and Technology (KAIST), Daejeon 34141, Republic of Korea

[2] KAIST InnoCORE PRISM-AI Center, Korea Advanced Institute of Science and Technology (KAIST), Daejeon 34141, Republic of Korea

**Abstract:** Ultrasound-based elasticity imaging is a non-invasive technique for estimating tissue stiffness fields from displacement fields obtained by comparing ultrasound signals before and after compression. While recent deep learning approaches have enabled faster and more accurate elasticity estimation compared to traditional methods, several challenges remain for clinical translation. In this study, we employ finite element simulations of free-hand palpation to investigate the applicability of the Fourier neural operator (FNO). Four practical scenarios were investigated: (1) prediction across diverse lesion morphologies, (2) generalization to cases with lesion counts differing from those in the training data, (3) robustness to noise in measured displacement fields, and (4) resilience to variations in ultrasound device resolution. Across these tasks, FNO consistently outperformed baseline models such as U-Net and DeepONet in predictive accuracy and generalization, while maintaining robustness under noise and resolution changes. Validated through *in silico* simulations, these findings demonstrate the potential of FNO as a framework that could facilitate translation of elasticity imaging toward clinical practice.

**Keywords:** elasticity imaging, Fourier neural operator, operator learning, deep learning, material parameters

[*] Corresponding author: ryush@kaist.ac.kr

# 1 INTRODUCTION

Elasticity imaging is a non-invasive technique for characterizing the material properties of biological tissues, thereby enabling differentiation between healthy and pathological states. For example, liver fibrosis is associated with increased tissue stiffness [1], while in breast cancer the spatial heterogeneity of the shear or elastic modulus is clinically relevant [2–4]. Among the available modalities—such as ultrasound, magnetic resonance elastography, and digital image correlation—ultrasound-based elasticity imaging has gained particular attention owing to its cost-effectiveness, real-time capability, and natural presence of speckle artifacts that serve as markers for displacement estimation [5].

In ultrasound elasticity imaging, tissue displacement is estimated by analyzing the speckle patterns of the tissue before and after deformation caused by an external force. Elasticity is then inferred from the measured displacement field using physics-based models, which are generally categorized into direct and iterative approaches. Direct methods simplify the stress equilibrium equation under assumptions such as incompressibility or local homogeneity, allowing direct computation of the elastic modulus field [6–8]. While computationally efficient, they require smooth strain fields and are highly sensitive to noise. Iterative methods, in contrast, rely on the numerical solutions, *e.g.*, by finite element method (FEM) [9–11]. These approaches construct a sensitivity matrix describing the relationship between local elasticity variations and resulting perturbations in displacement field, and iteratively update the modulus distribution until the FEM-predicted displacement matches the measured field. Although often more accurate, they demand heavy computation and multiple iterations, rendering real-time inference impractical. Moreover, elasticity must be re-estimated for each patients, and noise in the displacement field critically amplifies errors [12]. Finally, a common limitation of both methods is that accurate solutions require both lateral and axial displacement fields (displacement in *x*- and *y*-direction, respectively, see **Figure 1a**); however, in ultrasound imaging the lateral displacement field has a resolution 3–5 times lower than that of the axial field, leading to inaccurate estimates [13]. Additionally, abrupt stiffness changes, such as hard inclusions, poses difficulty in applying the traditional methods [14].

The advent of deep learning has enabled fast and accurate elasticity estimation, with inference times on the order of milliseconds once trained [15–19]. Several studies have proposed data-driven models that directly estimate elasticity from measured displacement fields [17,20–23]. For instance, Tuladhar et al. [20] generated FEM-based datasets of axial displacements and elastic modulus fields with inclusions of arbitrary shapes, and demonstrated that convolutional neural networks such as U-Net [24] can achieve highly accurate reconstruction of elastic modulus distribution (mean absolute percentage error (MAPE) of 1.14%). A subsequent study [21] expanded the training dataset by diversifying boundary conditions by sampling regions of interest from larger domains, further improving predictive accuracy to MAPE around 0.3%. Moreover, models trained solely on simulation data have shown promising transfferability to clinical datasets, demonstrating their potential to capture complex anatomical structures. However, their predictive performance degraded significantly under introduction of Gaussian noise, particularly when the signal-to-noise ratio in the strain field decreases (up to MAPE of 56.3%), suggesting limited robustness to realistic experimental perturbations.

Despite recent progress, several obstacles impede the clinical transition of deep learning-based elasticity imaging. First, the number of lesions within a region of interest is typically unknown, raising the question of whether models can generalize across varying lesion counts. In general, deep learning models are known to achieve accurate predictions when the input closely resembles the training data, but they may fail to exhibit robustness when confronted with different types of data. Second, noise is unavoidable in ultrasound displace measurements [25], and commonly used architectures (*e.g.*, U-Net) are vulnerable to noisy inputs. Third, for a trained model to be broadly applicable in practice, it must maintain predictive accuracy even when the resolution of the input data changes. Conventional deep learning models (*e.g.*, fully connected layer and U-

Net), however, can typically perform inference only at the same resolution as the training data. Since ultrasound devices differ in resolution and clinical data are difficult to obtain, acquiring new training datasets and retraining models for each device is impractical. These challenges motivates architectures with strong out-of-distribution generalization, and noise and resolution robustness.

Physics-informed neural networks (PINNs) [26] and neural operators [27] have emerged as promising alternatives. Several studies have applied PINNs to elasticity imaging [28–32], in which physical laws (*e.g.*, boundary conditions and the momentum balance equation) are explicitly incorporated into the loss function, and train network to satisfy these constraints. Althrough computationally less demanding than FEM-based methods, they still rely on iterative optimization, leading to long runtimes (*e.g.*, 579 seconds for a $64 \times 64$ domain with hard inclusion on an NVIDIA Tesla V100 GPU [30]) that preclude real-time use. Neural operators, by contrast, learn mappings between function spaces, such as from displacement to elasticity fields, based on the universal approximation theorem for operators [33]. Once trained, they produce solutions rapidly and demonstrate strong generalization performances [34]. Among neural operators, the deep operator network (DeepONet) [27] and Fourier neural operator (FNO) [35] have shown success in diverse inverse problems, including welding, electrical impedance tomography, and full-waveform inversion [36–38]. The FNO is particularly attractive owing to its learning mechanisms leveraging low-frequency modes enhances robustness to high-frequency noise [35,39]. Furthermore, because the parameters of FNO are learned directly in Fourier space and functions in the physical domain are resolved by projection onto Fourier basis defined over the entire domain, the method is inherently adaptable to changes in the resolution of the data.

Building on these characteristics, this study investigates the applicability of the FNO to elasticity imaging by addressing four practical scenarios *in silico* (**Figure 1a**). Specifically, we examine (**Figure 1b**): (1) predictive and generalization performance across diverse lesion morphologies, (2) robustness to variations in lesion count between training and testing, (3) resilience to noisy displacement data, and (4) robustness to variations in input resolution. Together, these scenarios reflect key challenges encountered in clinical settings. To evaluate them systematically, all assessments were performed using FEM-generated datasets. First, the predictive and generalization performance of FNO is benchmarked against U-Net and DeepONet across Gaussian inclusions, hard inclusions, and realistic tumor shapes (**Section 3.1**). Second, the generalization capability of FNO is tested with datasets containing different lesion counts than those used in training (**Section 3.2**). Third, robustness to noise is assessed by introducing perturbations into displacement fields and comparing FNO with U-Net (**Section 3.3**). Finally, FNO trained on low-resolution data was evaluated on high-resolution inputs to verify its resolution invariance (**Section 3.4**).

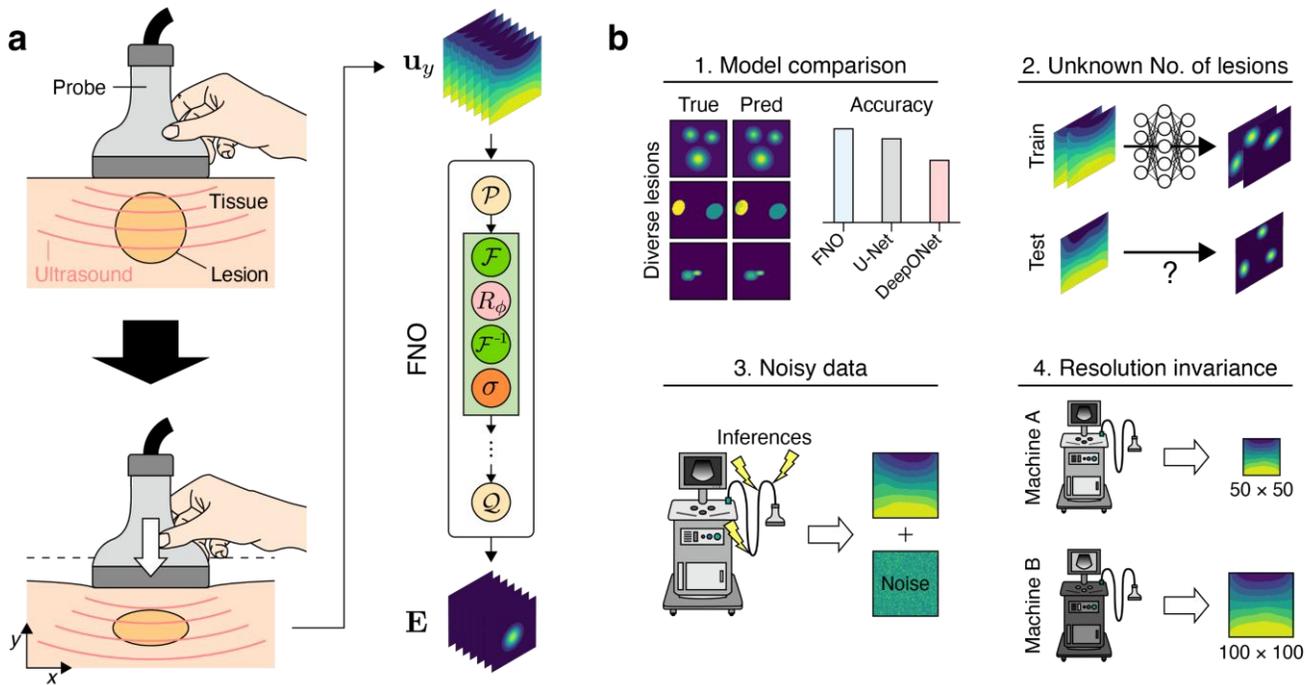

**Figure 1.** Schematic overview of the study. (a) In ultrasound-based elasticity imaging, the elastic modulus field is estimated from displacement fields obtained by comparing speckle patterns before and after quasi-static compression. In this work, the mapping from displacement fields to modulus fields is performed using the FNO. (b) To assess the practicality of FNO, its performance is benchmarked against baseline models across diverse lesion morphologies, and further evaluated under scenarios involving mismatched lesion counts between training and testing, the presence of noise in measured displacement data, and variations in input resolution.

## 2 METHODS

### 2.1 Data Acquisition via FEM

The dataset was generated using the commercial finite element solver ABAQUS/Standard, following the modeling framework of Kallel et al. [40]. The computational domain was a square of $100 \times 100$ mm$^2$ containing lesions (**Figure 2a**). Boundary conditions were defined as follows: the displacement within bottom surface was fully constrained (encastre), the top surface was restricted in the $x$-direction to prevent slip, and a uniform pressure of 210 Pa was applied on the top surface. For the finite element analysis, we modeled the tissue as a linear elastic material. This is a common assumption for biological tissues under small static and quasi-static deformation [41]. A two-dimensional plane strain condition was assumed using CPE4R elements, with the Poisson's ratio fixed at 0.495 to account for near-incompressibility of soft tissues [42]. A mesh convergence analysis (**Supplementary Section S1.1**) confirmed that a mesh with $100 \times 100$ elements provided sufficient accuracy.

To evaluate the capability of the FNO to generalize across lesion types, three lesion morphologies were considered: Gaussian inclusion, hard inclusion, and realistic tumor shapes (**Figure 2b**). The spatially varying elastic modulus field for each case was assigned to FE analysis utilizing the user programming capability of ABAQUS, with user-defined field (USDFLD) subroutine [43]. Each lesion morphologies is characterized as follows:

- **Gaussian inclusions**: The modulus was maximal at the center and decayed smoothly toward the periphery. The center location, length of major and minor axes, angle of major axis relative to the $x$-axis, and peak modulus were randomized.
- **Hard inclusions**: Elliptical inclusions with sharp modulus discontinuities were modeled, representing cases that violate the assumption of smooth material property fields and are challenging for direct methods. Geometric parameters and modulus values were randomized as in the Gaussian case.

- **Realistic tumors**: Shapes mimicking benign and malignant breast tumors were generated, reflecting the homogeneous stiffness of benign tumors and the heterogeneous stiffness of malignant ones [44–47]

For Gaussian and hard inclusion cases, the number of lesions ranged from one to three, with 2,000 paired datasets of axial displacement and modulus fields generated for each lesion count. For realistic tumors, a single lesion was modeled, and 2,000 datasets were generated. Further details of lesion modeling are provided in **Supplementary Section S1.2**.

Meanwhile, to enable the FNO to learn the operator that maps the axial displacement to the elastic modulus, displacement and modulus values should be obtained at the same spatial locations. In FEM, however, displacements are stored at nodal points, while elastic modulus are stored at element integration points (see **Figure 2a**). To align them, nodal displacements were interpolated to the integration points.

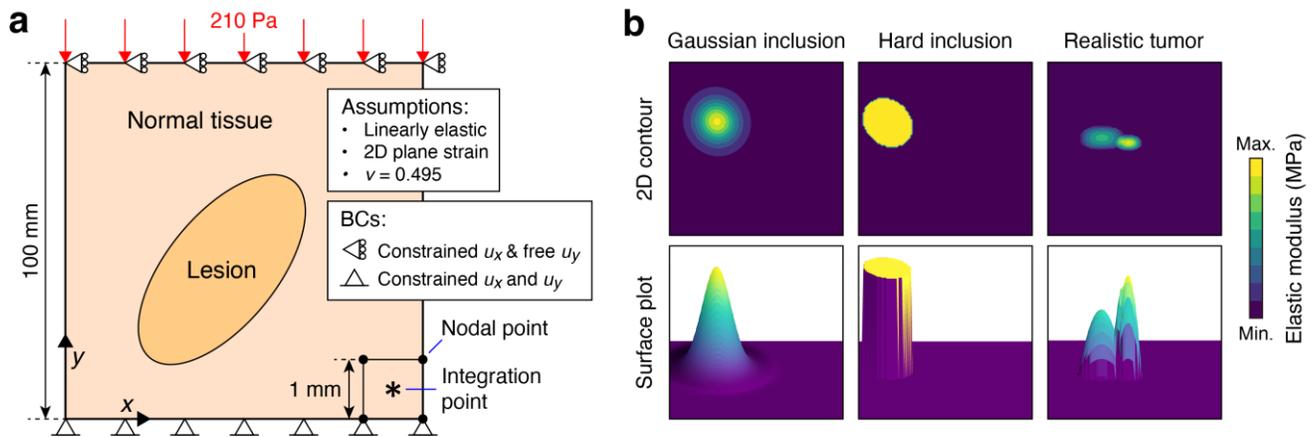

**Figure 2.** (a) Modeling assumptions employed in FEM for the data construction. (b) Contour and surface plots of elastic modulus fields for Gaussian inclusion, hard inclusion, and realistic tumor cases.

## 2.2. Fourier Neural Operator

### 2.1.1 Theoretical Background

Let the domain $D \subset \mathbb{R}^2$ be a bounded open set in two-dimensional Euclidean space. Define $\mathcal{U}(D; \mathbb{R}^{d_u})$ and $\mathcal{E}(D; \mathbb{R}^{d_E})$ as separable Banach spaces of functions taking values in $\mathbb{R}^{d_u}$ and $\mathbb{R}^{d_E}$, respectively, where $d_u = d_E = 1$ correspond to the dimensionality of scalar axial displacement and elastic modulus fields.

The goal of a neural operator is to approximate a target operator $\mathcal{G} : \mathcal{U} \mapsto \mathcal{E}$ from discrete observations $\{u(x_i, y_j), E(x_i, y_j)\}_{i,j=1}^{N}$ (following coordinate system defined in **Figure 2**a). The trained neural operator $\mathcal{G}_\phi$ thus learns a functional mapping from the displacement field to the elastic modulus:

$$\mathcal{G}_\phi : \mathcal{U} \ni u \mapsto E \in \mathcal{E}$$

where $\phi$ denotes the trainable network parameters.

An FNO comprises three major components (**Figure 3**). First, the input field $u(\mathbf{x})$ is projected into a higher-dimension latent space through a lifting layer $\mathcal{P}$, producing feature map $\mathbf{v}_0 = \mathcal{P}(u(\mathbf{x}))$. This operation enhances representational capacity and enables the network to learn nonlinear mappings between functional spaces. The latent feature is then updated iteratively through the rule:

$$\mathbf{v}_{t+1}(\mathbf{x}) = \sigma( \underbrace{W \mathbf{v}_t(\mathbf{x})}_{\text{real path}} + \underbrace{(\mathcal{K}(u; \phi) \mathbf{v}_t)(\mathbf{x})}_{\text{frequency path}} ), \quad t = 0, \dots, T-1, \tag{1}$$

where $W$ is a linear transformation, $\sigma$ is a nonlinear activation, and $\mathcal{K}(u; \phi)$ is a kernel integral operator parametrized by a neural network with learnable parameter $\phi$:

$$(\mathcal{K}(u; \phi) \mathbf{v}_t)(\mathbf{x}) = \int_D \kappa_\phi(\mathbf{x}, \xi, \mathbf{u}(\mathbf{x}), \mathbf{u}(\xi)) \mathbf{v}_t(\xi) \, d\xi. \tag{2}$$

where $\kappa_\phi$ is a neural network-based kernel function and $\xi$ is integrand variable. Conceptually, each Fourier layer (Eq. (1)) follows the standard neural-network pattern—a linear map followed by a nonlinear activation—to update hidden features, enabling the neural operator to represent nonlinear operators. The kernel integral operator (Eq. (2)) is inspired by Green's functions, the solution kernels of linear PDEs. When the kernel depends only on relative position, $\kappa_\phi(\mathbf{x}, \boldsymbol{\xi}) = \kappa_\phi(\mathbf{x} - \boldsymbol{\xi})$, the operator reduces to a convolution, directly paralleling the Green's-function representation of linear PDEs. Thus, while grounded in the solution structure of linear PDEs, the architecture is expressly designed to learn more complex, nonlinear operators.

Different parametrization of $\mathcal{K}$ yield different neural operator formulations. For example, DeepONet discretizes Eq. (2) pointwise, whereas the FNO exploits the convolution-Fourier duality:

$$(\mathcal{K}(u;\phi)\,\mathbf{v}_t)(\mathbf{x}) = \mathcal{F}^{-1}(R_\phi \cdot \mathcal{F}(\mathbf{v}_t))(\mathbf{x}), \tag{3}$$

where $\mathcal{F}$ and $\mathcal{F}^{-1}$ denote the Fourier and inverse Fourier transforms, respectively, and $R_\phi = \mathcal{F}\kappa_\phi$ is a learnable spectral filter. In practice, the Fourier series is truncated to retain a finite number of modes $k_{\max}$, leading to the discrete update rule:

$$\mathbf{v}_{t+1}(\mathbf{x}) = \sigma(\underbrace{W\mathbf{v}_t(\mathbf{x})}_{\text{real path}} + \underbrace{\mathcal{F}^{-1}(R_\phi \cdot \mathcal{F}(\mathbf{v}_t))}_{\text{frequency path}}), \quad t = 0, \dots, T-1. \tag{4}$$

By retaining $k_{\max}$ low-frequency modes in $R_\phi$, the model concentrates on the global, structural content of the solution and tends to be robust to high-frequency noise. As seen in Eq. (4), since $R_\phi$ is learned in Fourier space and $\mathcal{F}^{-1}(R_\phi \cdot \mathcal{F}(\mathbf{v}_t))$ amounts to projection onto Fourier basis functions defined over entire physical domain, FNO is robust to changes in input resolution. After $T$ iterations, the final representation $\mathbf{v}_T$ is mapped to the target field through a projection layer $\mathcal{Q}$.

### 2.1.2 FNO Implementation

The implementation follows the formulation in **Section 2.1.1** and the reference code in [35], with model hyperparameters and training configurations summarized in **Table 1**. The input tensor consists of the measured displacement field concatenated with spatial coordinates, forming a three-channel tensor. A lifting layer transforms it into a latent representation $\mathbf{v}_0 \in \mathbb{R}^{d_{v_0}}$, where $d_{v_t}$ is the $t$-th hidden dimension, listed in **Table 1**.

Each spectral convolutional layer contains two parallel branches (see **Figure 3**): (i) a frequency path, where the feature map is transformed to the Fourier domain via fast Fourier transform (FFT), truncated to the lowest $k_{\max} = 12$ modes, multiplied by the learable weights $R_\phi$, and transformed back via inverse FFT, and (ii) a real path, where the feature map undergoes a 1D convolution $W\mathbf{v}_t(\mathbf{x})$. The two branches are summed and passed through a ReLU activation (Eq. (4)). After four iterations ($T = 4$), the resulting latent field is projected through $Q$ to produce the predicted modulus field. Further details on the FNO implementation can be found in **Supplementary Section S2.1**.

Normalization of the inputs was performed via z-score normalization. The loss function was defined as the normalized L2 error:

$$L = \sum_{k=1}^{B} \frac{\left\| \mathbf{E}_k - \hat{\mathbf{E}}_k \right\|_2}{\left\| \hat{\mathbf{E}}_k \right\|_2}, \tag{5}$$

where $\mathbf{E}_k$ and $\hat{\mathbf{E}}_k$ denote the ground-truth and predicted elasticity fields for the $k$-th batch sample, $\|\cdot\|_2$ is the L2-norm, and $B$ is the batch size.

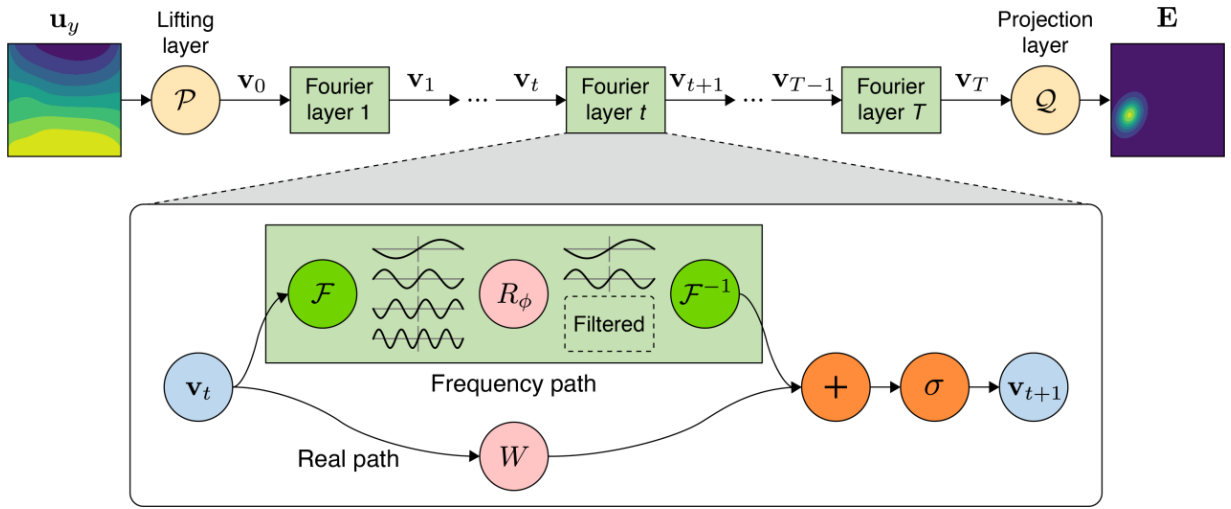

**Figure 3.** Model architecture of the FNO. The input displacement field is lifted into a high-dimensional representation, updated iteratively in Fourier space with nonlinear activations, and finally projected to the target elastic modulus field.

**Table 1.** Hyperparameters and training configurations of the FNO model used in this study.

| **Hyperparameters** | **Value** |
|---|---|
| Maximum No. of modes ($k_{\max}$) | 12 |
| No. of iterative architectures ($T$) | 4 |
| Hidden dimension ($d_{v_t}$, $t = 0, \dots, T-1$) | [128, 128, 128, 128] |
| Normalizer | Gaussian normalizer |
| Loss function ($L$) | Eq. 5 |
| Optimizer | Adam (`weight_decay = 1e-4`) |
| Learning rate | 1e-3 |
| Learning rate scheduler | StepLR (`step_size = 100`, `gamma = 0.5`) |
| Batch size ($B$) | 32 |
| Epochs | 1,000 |

## 3. RESULTS AND DISCUSSION

To demonstrate the effectiveness of the FNO in elasticity imaging, we evaluated its performance on four tasks: (1) prediction across diverse lesion morphologies, (2) generalization to an unknown number of lesions, (3) robustness to noisy inputs, and (4) robustness to variations in input resolution. Model accuracy was quantified using the mean relative error (MRE), which provides a scale-independent measure of discrepancy and avoids bias from variations in the magnitude of target values (note absolute error measures such as mean absolute error or root mean square error are sensitive to the scale of target values). The MRE is defined as

$$\text{MRE (\%)} = 100 \times \frac{1}{N^2} \sum_{j=1}^{N} \sum_{i=1}^{N} \left| \frac{E(x_i, y_j) - \hat{E}(x_i, y_j)}{E(x_i, y_j)} \right|,$$

where $E(x_i, y_j)$ and $\hat{E}(x_i, y_j)$ denote the ground truth and predicted elastic modulus at the $(x_i, y_j)$ grid point, and $N = 100$ is the total number of grid points along each coordinate direction ($x$ and $y$) in the $100 \times 100$ domain (see **Figure 2a**).

## 3.1. Generalization to Diverse Lesion Types

To evaluate the generalization performance of the FNO, we assessed its predictive capability on three lesion types: Gaussian inclusion, hard inclusion, and realistic tumor shapes—as well as on mixed datasets containing all three. Performance was benchmarked against two baseline models, U-Net and DeepONet, whose model architectures are described in **Supplementary Sections S2.2** and **S2.3**, respectively. For a fair comparison, all models were configured with the same normalizer, optimizer, learning rate, learning rate scheduler, batch size, and loss function.

For the Gaussian and hard inclusion datasets, cases with one, two, or three lesions were considered, while realistic tumor considered single lesion. Each dataset comprises 2,000 data, and was split 9:1 into training and test sets, yielding 1,800 training samples and 200 test samples per case. The real tumor dataset was divided in the same manner. For the mixed dataset, we constructed 3,500 training samples by randomly selecting 500 data from each of the seven cases (one to three Gaussian lesions, one to three hard inclusions, and realistic tumors), and 420 test samples by selecting 60 from the remaining data for each case. During training, 10% of the training data were reserved for validation. Training was terminated if the validation loss failed to improve by more than 0.0001 over 100 epochs (early stopping), and the model weights corresponding to the lowest validation loss were used for evaluation. Predictive performance was quantifies using the averaged MRE, defined as the mean MRE across all test samples.

The quantitative results across all cases are summarized in **Table 2**, and a qualitative comparison between FNO predictions and the ground truth for single-lesion test data is shown in **Figure 4**. Additional FNO predictions for cases with two or three Gaussian or hard inclusions and for the mixed dataset are provided in **Supplementary Figures S4** and **S5**, respectively. Overall, FNO achieved the lowest averaged MRE in all cases except when the domain contained two or three hard inclusions (see **Table 2**). For those cases, U-Net achieved slightly lower errors, reflecting stronger performance in capturing sharp modulus discontinuities. In contrast, DeepONet consistently produced higher errors than both U-Net and FNO, particularly for hard inclusions, where it tended to approximate abrupt boundaries with smooth, Gaussian-like transitions—an inherent limitation of linear reconstruction-based approaches [48].

As the number of lesions increased, the performance of FNO gradually declined, which is consistent with expectations since multiple lesions introduce complex lesion-lesion interactions that complicate operator learning. Nonetheless, the model maintained relatively low reconstruction errors even for three Gaussian lesions (averaged MRE of 1.026%). On the other hand, hard inclusions led to systematically higher errors with averaged MRE reached 1.895% for three inclusions, particularly concentrated along lesion boundaries (see **Figure 4** and **Supplementary Figure S4**). This behavior is theoretically anticipated, as accurately representing abrupt spatial variations in elastic modulus requires a large number of high-frequency Fourier modes. Although nonlinear activation can partially reconstruct such high-frequency information from the limited spectral components retained by the network [35], the recoverable range remains inherently constrained. Recent studies have therefore explored hybrid strategies to enhance the representation by sharp discontinuities by combining FNO with U-Net architectures [49], or by replacing the Fourier basis with wavelet expansions, as in the wavelet neural operator [50]. Aside from these boundary regions, the FNO predictions exhibit low overall errors and strong qualitative agreement with the ground truth distributions.

For realistic tumor shapes, localized errors were observed near lesion boundaries, but overall predictive accuracy remained high, with an averaged MRE of 0.193%. Finally, for the mixed dataset (**Supplementary Figure S5**), the averaged MRE was 0.352%, confirming that FNO can effectively learn across heterogeneous lesion morphologies. Collectively, these results highlight the strong predictive capability of FNO and suggest its promise for reliable performance when extended to real clinical data.

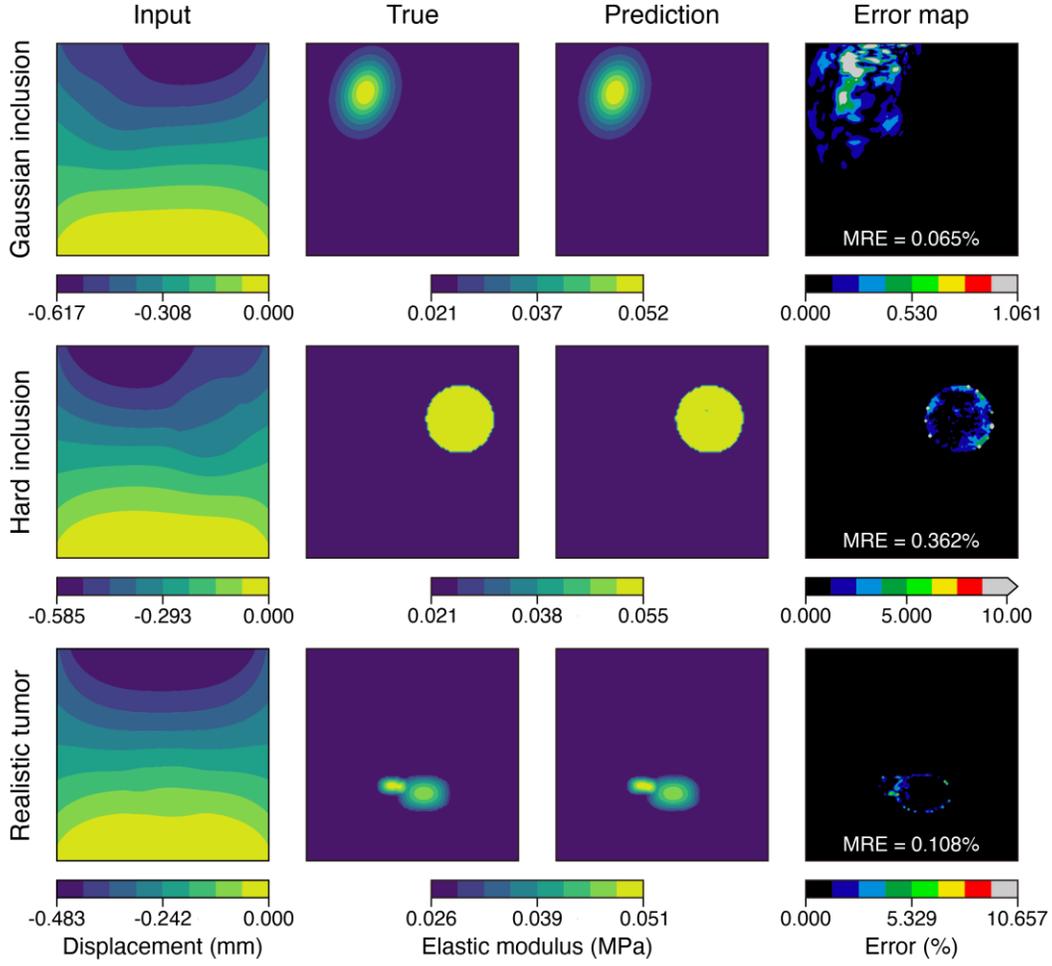

**Figure 4.** Predicted elastic modulus fields from the FNO and the corresponding ground truth, together with error maps, for representative lesion types given the axial displacement field. For clearer comparison, the colorbar ranges of the "True" and "Prediction" fields are matched.

**Table 2.** Quantitative evaluation of FNO, U-Net, and DeepONet on test data for three lesion types (red: best case).

|  |  | **Gaussian inclusion** | | | **Hard inclusion** | | | **Realistic tumor** | **Mixed data** |
|---|---|---|---|---|---|---|---|---|---|
| **No. of lesions** | | 1 | 2 | 3 | 1 | 2 | 3 | - | - |
| Averaged MRE (%) | FNO | 0.101 | 0.529 | 1.026 | 0.446 | 1.290 | 1.895 | 0.193 | 0.352 |
|  | U-Net | 0.250 | 0.809 | 1.077 | 0.514 | 1.038 | 1.892 | 0.341 | 0.858 |
|  | DeepONet | 1.404 | 2.250 | 3.579 | 5.991 | 10.01 | 11.04 | 0.770 | 4.983 |

## 3.2. Unknown Number of Lesions

In clinical practice, the number of lesions within a region of interest is often unknown in elasticity imaging. As a result, the number of lesions in training and test data may not match. Since deep learning models generally suffer performance degradation when confronted with out-of-distribution data [51–53], this scenario may represent a vulnerability for FNO. We therefore examined whether FNO can maintain reliable predictions when lesion counts differ between training and testing.

Two cases were considered with the lesion type fixed to Gaussian inclusion. In Case 1, the model was trained on datasets containing one and two lesions and tested on data with three lesions. In Case 2, the model was trained on single-lesion data and tested on data with two and three lesions. For Case 1, 1,000 samples were randomly selected from both the single- and double-lesion datasets to construct a training set of 2,000 samples, while 400 samples were drawn from the triple-lesion dataset for testing. For Case 2, all 2,000 single-lesion samples were used for training, and 200 samples each from the double- and triple-lesion datasets were seleted or testing (400 in total).

The results are shown in **Figure 5** and **Table 3**. In Case 1, FNO maintained robust predictive performance with an averaged MRE of 1.840%, even through the test set contained more lesions than seen in training. This robustness reflects the fact that neural operators learn the mapping from displacement fields to modulus fields rather than memorizing specific examples. In contrast, in Case 2 the model sometimes predicted only a single lesion when multiple lesions were present, and the averaged MRE increased to 4.829%. This degradation is likely due to the increased complexity of stress interactions among multiple lesions [54], which the model could not capture properly when trained solely on single-lesion data. These results suggest that incorporating diverse lesion counts during training enables FNO to generalize more effectively and maintain predictive accuracy in scenarios with previously unseen number of lesions.

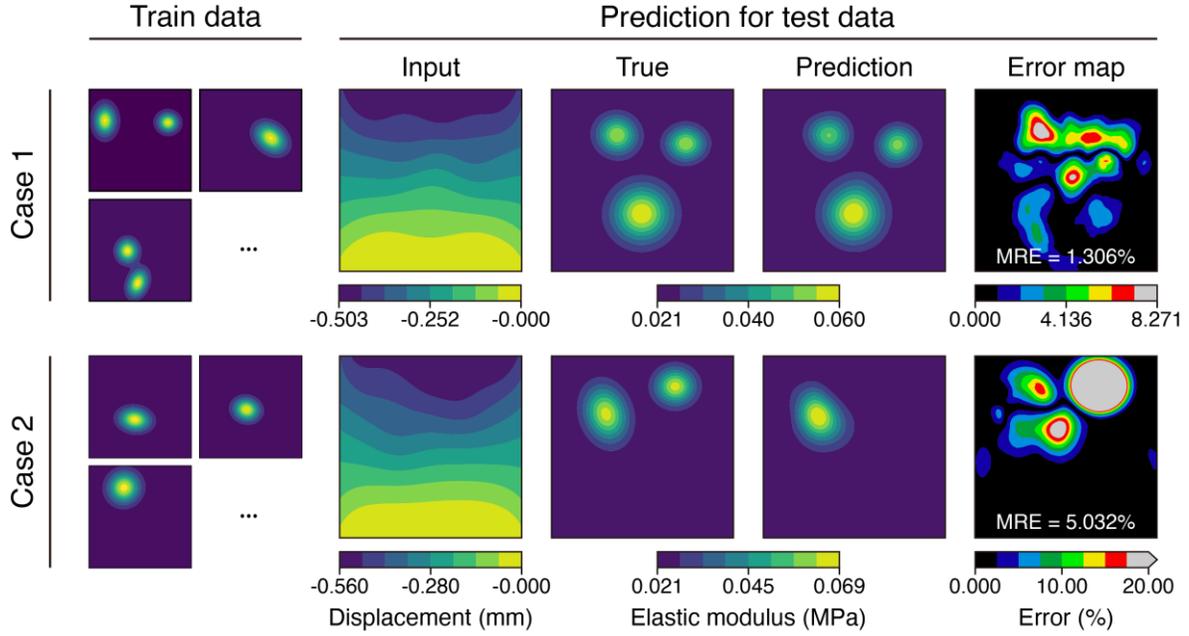

**Figure 5.** Validation of FNO for cases where the number of lesions in the training and test data differ. Case 1: training with one and two lesions, testing with three. Case 2: training with one lesion, testing with two and three. For clearer visual comparison, the colorbar ranges of the "True" and "Prediction" were matched.

**Table 3.** FNO model performance on test data when the number of lesions in the training and test dataset differ.

|  | **Case 1** | **Case 2** |
|---|---|---|
| Averaged MRE (%) | 1.840 | 4.829 |

### 3.3. Robustness to the Noisy Input Data

In ultrasound-based tissue displacement estimation, electronic and acoustic interferences introduce noise into the measured signals, which can degrade the performance of deep learning models. Prior studies have shown that models trained on noise-free data often deteriorate when tested on noisy inputs [55]. In clinical practice, noise levels may also vary across devices, raising the questions of whether an FNO trained on one dataset can remain robust when applied to data collected under different noise conditions. To evaluate this, we tested one of the most challenging scenarios: training on noise-free displacement fields and testing on noisy ones.

The experiment was conducted using Gaussian inclusions, with the data split 9:1 into training and test sets. Noisy test data, $\tilde{u}$, were generated by adding zero-mean Gaussian noise to the displacement fields:

$$\tilde{u}(x_i, y_j) = u(x_i, y_j) + \mathcal{N}\left(0, \alpha \times u(x_i, y_j)\right), \quad i,j = 1,2,\ldots,N,$$

where $u(x_i, y_j)$ denotes the axial displacement at the grid point $(x_i, y_j)$, $\alpha$ is a scale factor controlling the noise level, $\mathcal{N}(a, b)$ is the Gaussian distribution with mean $a$ and standard deviation $b$, and $N = 100$ is the total number of grid points along each coordinate direction (*x* and *y*). Performance of FNO and U-Net was compared for $\alpha$ values ranging from 0.01 to 0.05.

Qualitative results are shown in **Figure 6**, and quantitative outcomes in **Table 4**. As seen in the error maps of **Figure 6**, FNO consistently produced lower errors across all noise levels. U-Net predictions degraded rapidly with increasing noise, showing markedly distorted modulus maps from $\alpha = 0.02$ onward. In constrast, FNO predictions closely resembled the ground truth even at $\alpha = 0.05$. Quantitatively, FNO achieved an averaged MRE of 1.81% at $\alpha = 0.05$, whereas U-Net recorded 33.2% at $\alpha = 0.02$, increasing to 64.9% at $\alpha = 0.05$.

The robustness of FNO arises from its architecture. By restricting to 12 low-frequency Fourier modes, the FNO naturally filters out high-frequency noise, which is largely confined to discarded modes, while retaining low-frequency structural information that capture the tissue modulus distribution. This explains why FNO maintains reliable predictions under noise levels that severely degrade conventional deep learning models.

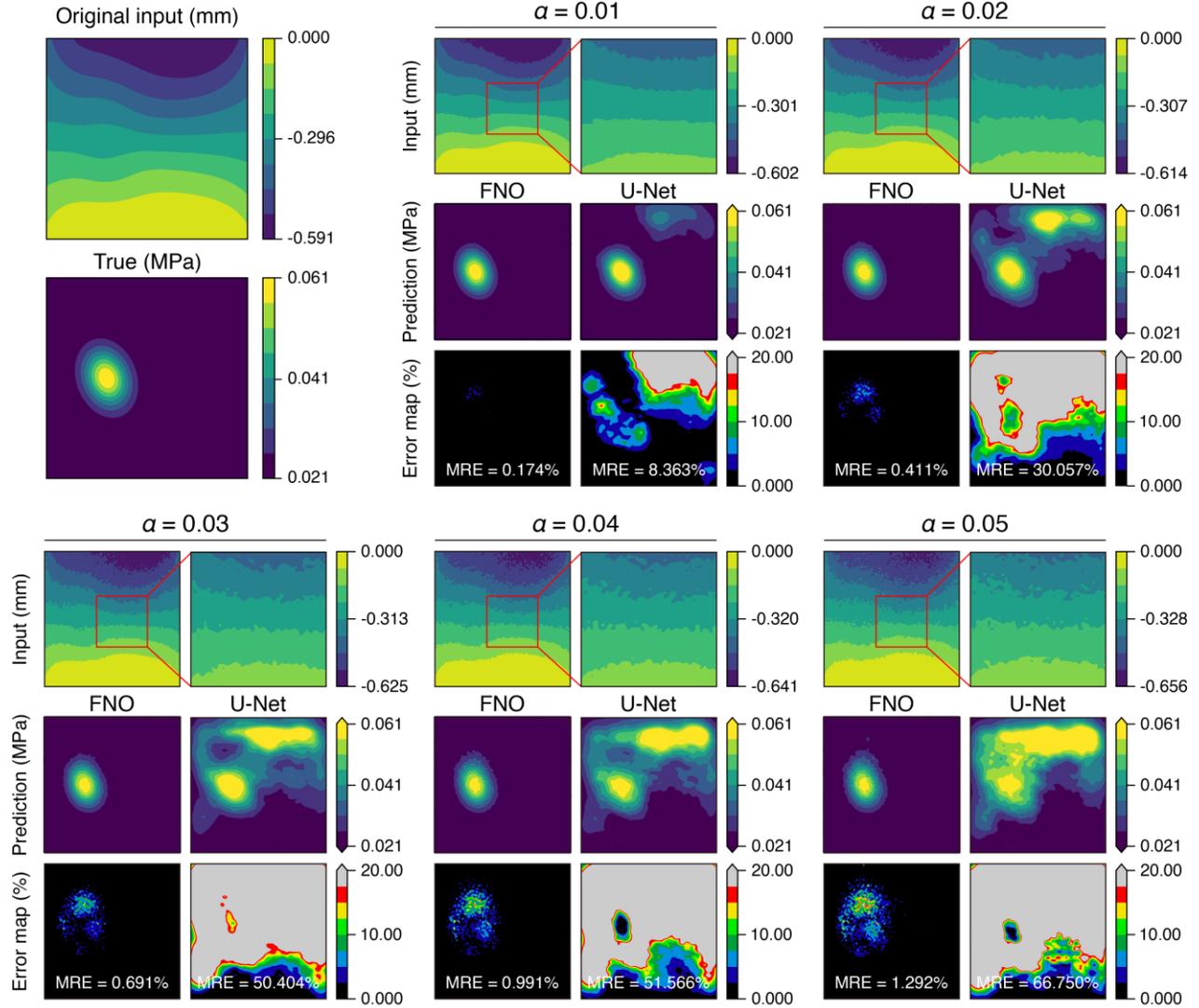

**Figure 6.** Predictions of FNO and U-Net for noisy input displacement fields. For clarity, the colorbar ranges of "True" and "Prediction" are aligned to that of "True". In the "Prediction" maps, values exceeding or falling below the ground truth are represented by the uppermost or lowermost colors. Error maps are aligned to the U-Net scale, and regions with errors greater than 20% are highlighted with the uppermost color.

**Table 4.** Prediction performance of FNO and U-Net on test data with noisy input displacement fields.

| | $\alpha$ | **0.01** | **0.02** | **0.03** | **0.04** | **0.05** |
|---|---|---|---|---|---|---|
| Averaged MRE (%) | U-Net | 9.10 | 33.2 | 51.9 | 52.1 | 64.9 |
| | FNO | 0.228 | 0.569 | 0.951 | 1.38 | 1.81 |

## 3.4. Unified Modeling Across Resolutions with FNO

Unlike conventional models such as fully connected networks or U-Net, which are restricted to fixed input and output dimensions, the FNO can predict the elastic modulus field across arbitrary input resolutions. This capability stems from its operator-learning formulation: as described in **Section 2.1.2**, the lifting layer $\mathcal{P}$, spectral convolution layers, and projection layer $\mathcal{Q}$ act independently of the discretization of the input domain. By performing learning in the spectral domain and reconstructing solutions through projection onto Fourier bases that span the entire field, the FNO inherently preserves consistency across different grid or sampling resolutions [35].

This property is particularly advantageous in clinical settings, where retraining separate models for imaging devices with varying resolutions is often impractical due to the high cost of new data acquisition. To examine this property, we evaluated the FNO using training and test datasets with mismatched resolutions. The test dataset was fixed at a resolution of $100 \times 100$, while training datasets were generated by subsampling the original field at intervals of two, three, and four pixels, yielding coarser resolutions of $50 \times 50$, $34 \times 34$, and $25 \times 25$, respectively. This setup assessed whether an FNO trained on lower-resolution data—representing acquisition from lower-cost devices—could still provide accurate predictions when applied to higher-resolution measurements obtained from more advanced systems.

**Figure 7** presents the qualitative results, and **Table 5** summarizes the quantitative evaluation. FNO trained on $50 \times 50$ (**Figure 7a**) and $34 \times 34$ (**Figure 7b**) datasets maintained strong predictive performance when tested on $100 \times 100$ data, with averaged MRE values of 1.391% and 1.185%, respectively. In contrast, training on the $25 \times 25$ dataset (**Figure 7c**) led to distorted predictions of Gaussian inclusions and a higher averaged MRE of 4.649%, reflecting a notable loss of accuracy. These findings indicate that FNO can provide reliable predictions even when the input resolution differs from that of the training data, achieving robust performance up to 8.65 times higher resolution (which is $100^2 / 34^2$). Overall, these results underscore the ability of FNO to bridge resolution gaps across devices, highlighting its potential as a practical tool for real-world elasticity imaging applications.

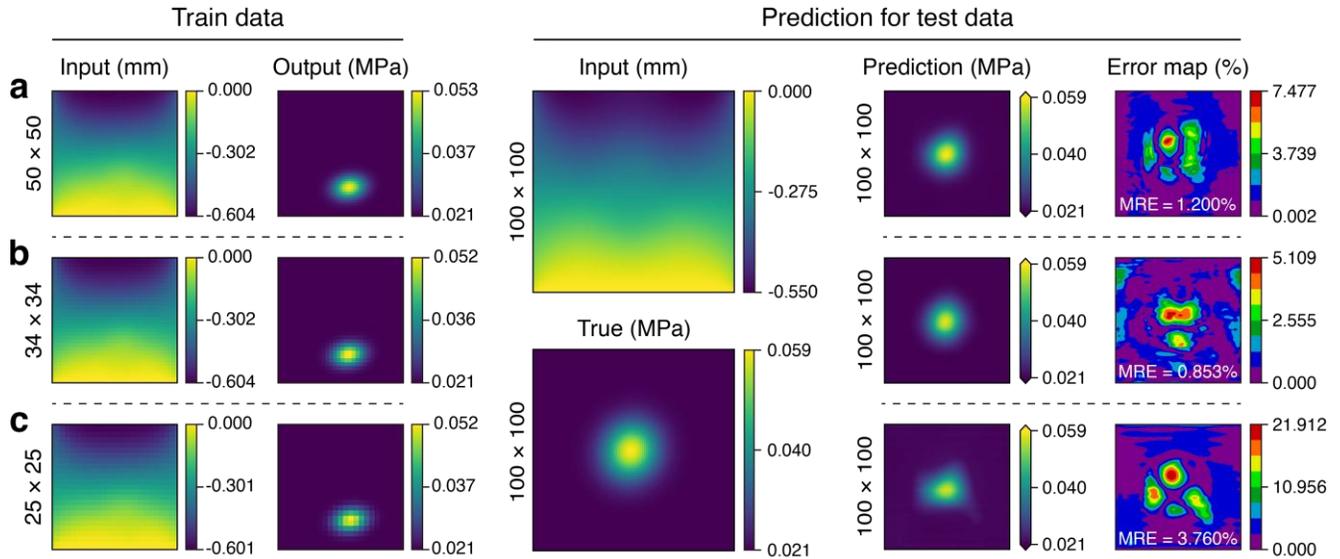

**Figure 7.** Predictive performance of FNO when the resolutions of the training and test datasets differ. (a) Training dataset resolution: 50 × 50; test datast resolution: 100 × 100. (b) Training dataset resolution: 34 × 34; test datast resolution:100 × 100. (c) Training dataset resolution: 25 × 25; test datast resolution: 100 × 100. For clearer visual comparison, the colorbar ranges of "True" and "Prediction" were matched to those of "True". In the predictions, "Input" and "True" fields are identical across (a), (b), and (c).

**Table 5.** Prediction performance of the FNO on 100 × 100 resolution test data for training data with different resolutions.

| Training dataset resolution | 50 × 50 | 34 × 34 | 25 × 25 |
|---|---|---|---|
| Averaged MRE (%) | 1.391 | 1.185 | 4.649 |

## 4. CONCLUSION

This study investigated the Fourier Neural Operator (FNO) as a robust operator-learning framework for elasticity imaging. Through systematic evaluations under four clinically relevant scenarios—diverse lesion morphologies, mismatched lesion counts, noisy displacement fields, and resolution variations—the results indicate that FNO consistently ourperformed conventional deep learning architectures, except in multi-lesion cases with hard inclusions, where U-Net achieved slightly lower errors.

The findings underscore three major advantages of the FNO. First, by learning an operator that maps axial displacement fields to elasticity distributions, FNO maintained predictive accuracy across unseen lesion configurations, demonstrating strong generalization capability. Second, its spectral formulation which is limited to a compact set of low-frequency modes, suppressed noise amplification and yielded significantly improved robustness compared with U-Net. Third, the decoupling of spectral operations from spatial discretization enabled resolution-invariant predictions, minimizing dependence on device-specific acquisition settings.

Nevertheless, truncation of high-frequency modes biases the representation toward global structures, leading to elevated errors near sharp modulus transitions. This limitation may be mitigated by hybrid architectures that combine FNO with convolutional decoders such as U-Net [49], or by replacing the Fourier basis with multi-scale representations such as wavelets [50]. Such extensions, however, must balance potential gains in local resolution against possible losses in computational efficiency and noise tolerance.

While this study was limited to simulation data, the results establish FNO as a theoretically grounded and practically robust approach that addresses key challenges of conventional deep learning models in elasticity imaging. Future research should extend this framework to clinical datasets and explore integration with physics-based regularization [56] or multi-fidelity [57] and transfer learning [58] strategies to further improve data efficiency and generalization.

**Acknowledgement:** This work was supported by the InnoCORE program of the Ministry of Science and ICT (N10250154).

**Data Availability Statement:** The used data are available on reasonable request.

**Conflicts of Interest:** The authors declare no conflict of interest.

Supplementary Materials for

# Morphology-, Noise-, and Resolution-Robust Ultrasound Elasticity Imaging with Fourier Neural Operators

Heekyu Kim[1], Hugon Lee[1], Minwoo Park[1], and Seunghwa Ryu[1,2*]

[1]Department of Mechanical Engineering, Korea Advanced Institute of Science and Technology

(KAIST), Daejeon 34141, Republic of Korea

[2]KAIST InnoCORE PRISM-AI Center, Korea Advanced Institute of Science and Technology

(KAIST), Daejeon 34141, Republic of Korea

## Contents

---
*Corresponding author. ryush@kaist.ac.kr

## S1    Data acquisition

### S1.1    FEM mesh convergence analysis

To verify the credibility of the FEM simulations (Figure 2a) performed throughout this study, a mesh convergence analysis was conducted. Since this study trains the FNO using displacement–elastic modulus field pairs, convergence was assessed based on the maximum displacement magnitude. The displacement magnitude was computed as $(u_x^2 + u_y^2)^{1/2}$ using both the lateral displacement $u_x$ and axial displacement $u_y$. Although only the axial displacement field was used for training the FNO, both displacement components were considered for this calculation to more comprehensively evaluate convergence. As shown in Figure S1, convergence behavior was observed when the element size reached approximately 3 mm, providing error under 1 % compared to 0.5 mm element size case. Nevertheless, to ensure precise discretization of the $100 \times 100$ mm$^2$ domain considered for this study and to conservatively secure data reliability, an element size of 1 mm was used throughout this study, where the relative error compared to element size of 0.5 mm case were listed in Table S1.

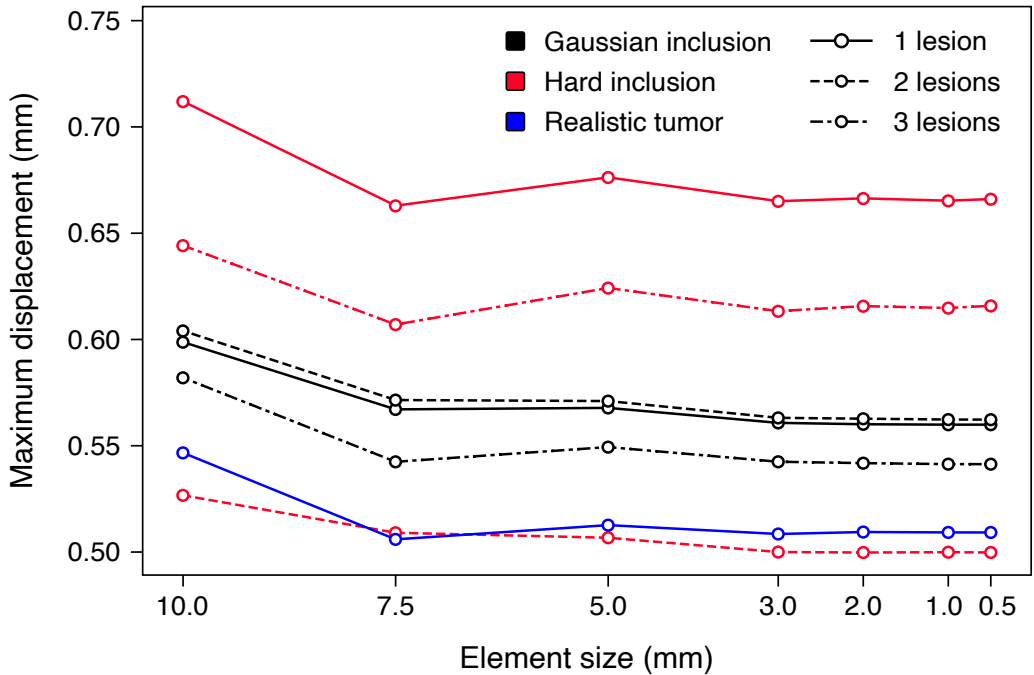

**Figure S1.** Result of mesh convergence analysis for various lesion types, for element sizes of 10.0, 7.5, 5.0, 3.0, 2.0, 1.0, and 0.5 mm.

**Table S1.** Relative error in the maximum displacement at an element size of 1.0 mm, referenced to the value corresponding to an element size of 0.5 mm in the convergence test.

|  | *Gaussian inclusion* | | | *Hard inclusion* | | | *Realistic tumor* |
|---|---|---|---|---|---|---|---|
| No. of lesions | 1 | 2 | 3 | 1 | 2 | 3 | - |
| Error ($10^{-2}$ %) | 0.203 | 0.902 | 0.213 | 11.5 | 2.46 | 16.8 | 0.4 |

**S1.2    Generation of elasticity fields**

This section describes how the elastic modulus fields were generated to construct the dataset, considering three types of scenarios: (1) Gaussian inclusions, (2) hard inclusions, and (3) realistic tumors. It is noted that symbol $\mathcal{U}(a,b)$ is assigned here for continuous uniform distribution within interval with minimum $a$ and maximum $b$, while the symbol $\mathcal{U}$ is utilized to designate function space for displacement fields within the main text.

S1.2.1    Gaussian inclusion

The Gaussian inclusion is defined such that the elastic modulus follows Gaussian distribution in two-dimension, which decreases outward from the center, approaching 21 kPa. The center position, the lengths of the major and minor axes, the rotation angle with respect to the x-axis, and the elastic modulus at the center are randomly varied. To express spatial variation of the elastic modulus field, a field variable $0 \leq F(x,y) \leq 1$ is defined as follows:

$$F(x,y) = c \cdot \exp\left(-\left[\left(\frac{x'}{m}\right)^2 + \left(\frac{y'}{n}\right)^2\right]\right), \tag{S1}$$

where

$$x' = (x - x_1) \cdot \cos\theta + (y - y_1) \cdot \sin\theta \quad \text{for} \quad 0 \leq x \leq 100,\ 0 \leq y \leq 100,$$
$$y' = (x - x_1) \cdot (-\sin\theta) + (y - y_1) \cdot \cos\theta \quad \text{for} \quad 0 \leq x \leq 100,\ 0 \leq y \leq 100.$$

Here, $(x_1, y_1)$ denotes the center coordinates of the Gaussian-shaped lesion, and $\theta$ represents the angle of the major axis of the inclusion with respect to the $x$-axis. The lesion center $(x_1, y_1)$, rotation angle $\theta$, and the coefficients $c$, $m$, and $n$ are drawn from the uniform distribution $\mathcal{U}(\min, \max)$ as follows:

$$x_1,\ y_1 \sim \mathcal{U}(15.0, 85.0),$$
$$\theta \sim \mathcal{U}(0.0, \pi),$$
$$c \sim \mathcal{U}(0.7, 0.9),$$
$$m,\ n \sim \mathcal{U}(8.0, 16.0).$$

When the value of field variable $F(x,y)$ is 1, a value of elastic modulus $0.063 + \mathcal{U}(-0.01, 0.01)$ MPa is assigned, whereas when $F(x,y)$ equals 0, a value of elastic modulus 0.021 MPa is assigned. For values between 0 and 1, the elastic modulus is interpolated as follows:

$$E(x,y) = 0.021 + [\{0.063 + \mathcal{U}(-0.01, 0.01)\} - 0.021] \times F(x,y) \quad [\text{MPa}].$$

When the number of lesions is two or more, the same modeling procedure is applied, except that a constraint is imposed to ensure that the distance between lesion centers is greater than 20, thereby preventing lesions from being placed so close that they appear as a single lesion. For example, when the number of lesions is two, the field variable $F(x,y)$ is modeled as follows:

$$F(x,y) = c_1 \cdot \exp\left(-\left[\left(\frac{x'_1}{m_1}\right)^2 + \left(\frac{y'_1}{n_1}\right)^2\right]\right) + c_2 \cdot \exp\left(-\left[\left(\frac{x'_2}{m_2}\right)^2 + \left(\frac{y'_2}{n_2}\right)^2\right]\right), \tag{S2}$$

where

$$x'_i = (x - x_i) \cdot \cos \theta_i + (y - y_i) \cdot \sin \theta_i \quad \text{for} \quad 0 \leq x \leq 100, \ 0 \leq y \leq 100,$$

$$y'_i = (y - y_i) \cdot (-\sin \theta_i) + (x - x_i) \cdot \cos \theta_i \quad \text{for} \quad 0 \leq x \leq 100, \ 0 \leq y \leq 100,$$

$$\text{with} \quad \sqrt{(x_1 - x_2)^2 + (y_1 - y_2)^2} \geq 20,$$

where $i = 1, 2$ designating each lesion. The parameters $(x_i, y_i)$, $\theta_i$, $c_i$, $m_i$, and $n_i$ ($i = 1, 2$) are drawn from the same uniform distributions as in the single lesion case. The field variable $F(x, y)$ is then converted to the elastic modulus in the same manner as described above.

### S1.2.2 Hard inclusion

A hard inclusion is characterized by an elliptical region with a higher elastic modulus embedded within the matrix. Similar to the Gaussian inclusion, the center position, lengths of the major and minor axes, the rotation angle with respect to the $x$-axis, and the elastic modulus inside the inclusion are ramdomly varied. The field variable $F(x, y)$ is modeled as follows:

$$F(x, y) = \begin{cases} 1, & \text{if } \left(\frac{x'}{m}\right)^2 + \left(\frac{y'}{n}\right)^2 \leq 1.0, \\ 0, & \text{otherwise}, \end{cases} \tag{S3}$$

where

$$x' = (x - x_1) \cdot \cos \theta + (y - y_1) \cdot \sin \theta \quad \text{for} \quad 0 \leq x \leq 100, \ 0 \leq y \leq 100,$$

$$y' = (x - x_1) \cdot (-\sin \theta) + (y - y_1) \cdot \cos \theta \quad \text{for} \quad 0 \leq x \leq 100, \ 0 \leq y \leq 100.$$

Here, $(x_1, y_1)$ denotes the center coordinates of the hard inclusion, and $\theta$ represents the angle of the major axis with respect to the $x$-axis. The lesion center $(x_1, y_1)$, rotation angle $\theta$, and the coefficients $m$ and $n$ are drawn from the uniform distribution $\mathcal{U}$ as follows:

$$x_1, y_1 \sim \mathcal{U}(15.0, 85.0),$$
$$\theta \sim \mathcal{U}(0.0, \pi),$$
$$m, n \sim \mathcal{U}(8.0, 16.0).$$

When the value of field variable $F(x, y)$ is 1, a value of elastic modulus $0.063 \times \mathcal{U}(0.8, 1.2)$ MPa is assigned, whereas when $F(x, y)$ equals 0, a value of elastic modulus $0.021 \times \mathcal{U}(0.8, 1.2)$ MPa is assigned.

When the number of hard inclusions is two or more, a conservative constraint is imposed to completely prevent overlap by requiring that the distance between the centers of any two inclusions be greater than the sum of their major axes. In addition, the hard inclusions are assigned different elastic moduli. For example, when the number of inclusions is two, the field variable $F(x, y)$ is modeled as follows:

$$F(x, y) = \begin{cases} 1, & \text{if } \left(\frac{x'_1}{m_1}\right)^2 + \left(\frac{y'_1}{n_1}\right)^2 \leq 1.0, \\ c, & \text{if } \left(\frac{x'_2}{m_2}\right)^2 + \left(\frac{y'_2}{n_2}\right)^2 \leq 1.0, \\ 0, & \text{otherwise}, \end{cases} \tag{S4}$$

where
$$x'_i = (x - x_i) \cdot \cos\theta_i + (y - y_i) \cdot \sin\theta_i \quad \text{for} \quad 0 \leq x \leq 100, \ 0 \leq y \leq 100,$$
$$y'_i = (x - x_i) \cdot (-\sin\theta_i) + (y - y_i) \cdot \cos\theta_i \quad \text{for} \quad 0 \leq x \leq 100, \ 0 \leq y \leq 100,$$
$$\text{with} \quad \max(m_1, n_1) + \max(m_2, n_2) < \sqrt{(x_1 - x_2)^2 + (y_1 - y_2)^2},$$

where $i = 1, 2$ designating each lesion. The parameters $(x_i, y_i)$, $\theta_i$, $m_i$, and $n_i$ ($i = 1, 2$) are drawn from the same uniform distributions as in the single lesion case. The coefficient $c$ is drawn from the uniform distribution
$$c \sim \mathcal{U}(0.6, 1.0),$$
to ensure that the two inclusions have different elastic moduli. The field variable $F(x, y)$ is then converted to the elastic modulus in the same manner as described above.

### S1.2.3   Realistic tumor

In modeling realistic tumor lesions, we take into account that the degree of heterogeneity in the elastic modulus can serve as a factor for distinguishing malignant from benign breast cancer. The modeling approach follows Patel et al. (2019), where the degree of heterogeneity is represented using the distance between two tumors. The field variable $F(x, y)$ is defined as follows:

$$F(x, y) = \sum_{i=1}^{2} \chi_i \cdot \left[ c_i \left( 1 - \left( \frac{x - x_i}{m_i} \right)^2 \right) \left( 1 - \left( \frac{y - y_i}{n_i} \right)^2 \right) \right], \tag{S5}$$

where
$$\chi_i = \begin{cases} 1, & \text{if } \left( \frac{x - x_i}{m_i} \right)^2 + \left( \frac{y - y_i}{n_i} \right)^2 \leq 1, \\ 0, & \text{otherwise}, \end{cases}$$
$$c_1 \sim \mathcal{U}(0.45, 0.60), \qquad c_2 \sim \mathcal{U}(0.60, 0.80),$$
$$m_1 \sim \mathcal{U}(8.0, 15.0), \qquad n_1 \sim \mathcal{U}(6.0, 10.0),$$
$$m_2 \sim m_1 \times \mathcal{U}(0.45, 0.65), \quad n_2 \sim n_1 \times \mathcal{U}(0.5, 0.65).$$

For the *benign* case, the center coordinates of the two tumors $(x_1, y_1)$ and $(x_2, y_2)$ are drawn from the uniform distribution as follows:
$$x_1, \ y_1 \sim \mathcal{U}(25.0, 75.0),$$
$$x_2 = x_1 + \phi(x_1) \times e_x,$$
$$y_2 = y_1 + \phi(x_2) \times e_y,$$

where
$$\phi(\lambda) = \begin{cases} 1 & \text{if } 0 \leq \lambda \leq l/2, \\ -1 & \text{if } l/2 \leq \lambda \leq l, \end{cases}$$
$$e_x \sim m_1 \times \mathcal{U}(0.2, 0.3),$$
$$e_y \sim n_1 \times \mathcal{U}(0.05, 0.15),$$

and $l = 100$ mm is the length of domain in each direction.

For the *malignant* case, the center coordinates are obtained as:

$$x_1, y_1 \sim \mathcal{U}(25.0, 75.0),$$
$$x_2 = x_1 + \phi(x_1) \times (m_1 + m_2 - e_x),$$
$$y_2 = y_1 + \phi(x_2) \times e_y,$$

where $\phi(\lambda)$ follows same definition and

$$e_x \sim m_1 \times \mathcal{U}(0.35, 0.45),$$
$$e_y \sim n_1 \times \mathcal{U}(0.35, 0.45).$$

When the value of field variable $F(x, y)$ is 1, a value of elastic modulus $0.063 + \mathcal{U}(-0.01, 0.01)$ MPa is assigned, whereas when $F(x, y)$ equals 0, a value of elastic modulus $\mathcal{U}(0.025, 0.035)$ MPa is assigned. For intermediate values, the elastic modulus is determined by linear interpolation.

## S2   Model architectures

This section details the architectures of FNO and the baseline models (U-Net and DeepONet) for performance comparison. For a fair comparison, training hyperparameters for the models (including normalizer, optimizer, learning rate, learning rate scheculer, batch size, and loss function) were kept identical across all models (compare Table 1 in main text).

### S2.1   FNO

The FNO architecture is illustrated in Figure 3, with hyperparameters listed in Table 1 of the main text. The input consists of the displacement field concatenated with the spatial coordinates ($x$ and $y$), forming a three-channel tensor of shape $(B, H, w, 3)$, where $B$ denotes the batch size, $H$ the height, and $w$ the width. The input is first processed by a lifting layer composed of linear layers to produce a feature map of shape $(B, H, w, d_{v_0})$. Here, $d_{v_t}$ (for $t = 0, 1, 2, 3$) correspond to the hidden dimensions listed in Table 1, where $d_{v_t} = 128$ is utilized in this study. To perform spectral convolution, the feature map is permuted to $(B, d_{v_0}, H, w)$ and iteratively updated following Eq. (4) in the main text.

Each spectral convolutional block consists of two parallel branches: (i) a frequency-domain branch and (ii) a real-space branch. In the frequency path, the feature map is transformed into the Fourier domain ($\mathcal{F}\mathbf{v}_t$) using the FFT, yielding a tensor of shape $(B, d_{v_t}, H, w/2 + 1)$. Along the $H$-axis, only the lowest frequency components within the first $k_{\max}$ and the last $k_{\max}$ indices (*i.e.*, slices $[:, :, : k_{\max}, :]$ and $[:, :, -k_{\max} :, :]$) are retained, while all other values were set to zero. Along the $w$-axis, only the lowest $k_{\max}$ frequency components (*i.e.*, slice $[:, :, :, : k_{\max}]$) are preserved, with the remaining values zeroed out. The truncated feature map is multiplied by a learnable weight tensor $R_\phi$ of shape $(d_{v_t}, d_{v_{t+1}}, k_{\max}, k_{\max})$, and transformed back into the spatial domain by inverse FFT. After four iterative updates ($T = 4$), the resulting feature map of shape $(B, d_{v_4}, H, w)$ with $d_{v_4} = 128$ is passed through a projection layer to yield the predicted elastic modulus field with shape $(B, d_E, H, w)$ where $d_E = 1$.

In parallel, the real path reshapes the feature map to $(B, d_{v_t}, H \times w)$, processes it with a one-dimensional convolutional ($W\mathbf{v}_t(\mathbf{x})$), and reshapes it back to $(B, d_{v_t}, H, w)$. The outputs of the frequency and real-space branches are summed and passed through a ReLU activation $\sigma$, producing $\mathbf{v}_{t+1}$ of shape $(B, d_{v_{t+1}}, H, w)$. Following the recommendation of Li et al. (2020), we set the maximum number of mode $k_{\max} = 12$ and the number of iterative architectures $T = 4$.

### S2.2   U-Net

The architecture of the U-Net used in this study is illustrated in Figure S2, and detailed hyperparameters are provided in Table S2. Using the notation (batch size, channel, height, width), the U-Net takes an axial displacement field of shape $(B, 1, 100, 100)$ as input, where $B$ is the batch size. After passing through a convolutional layer, it is transformed into feature maps of shape $(B, 64, 100, 100)$. The encoder then applies four down-sampling steps, where the number of channels doubles at each step while the width and height of the feature maps are halved. After all down-sampling operations, the feature maps have shape $(B, 1024, 6, 6)$. These are subsequently upsampled and passed through four blocks, each consisting of two sequential convolutional layers followed by batch normalization (`BatchNorm`) and a rectified linear unit (`ReLU`). The resulting feature maps of shape $(B, 64, 100, 100)$ are finally mapped to the target elastic modulus field of shape $(B, 1, 100, 100)$ through a convolutional layer.

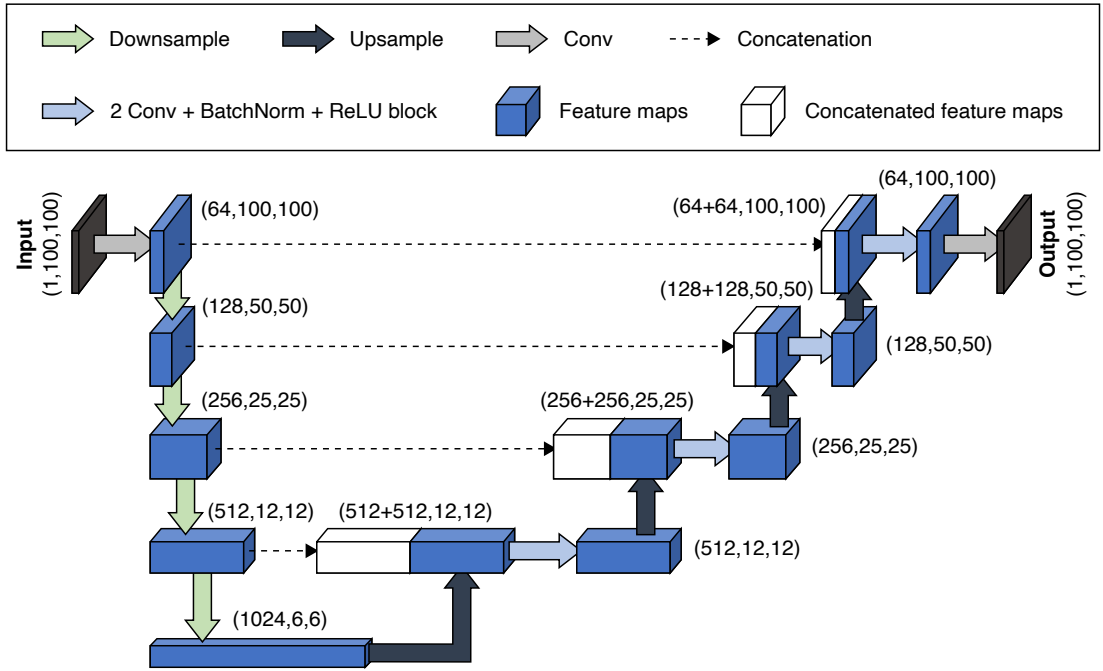

**Figure S2.** Architecture of the U-Net used in this study.

**Table S2.** Hyperparameters of the U-Net.

| Parameter | Value |
| --- | --- |
| Normalizer | Gaussian normalizer |
| Loss function | Eq. (5) (from the main text) |
| Optimizer | Adam (`weight_decay` $= 1 \times 10^{-4}$) |
| Learning rate | $1 \times 10^{-3}$ |
| Learing rate scheduler | StepLR (`step_size` $= 100$, `gamma` $= 0.5$) |
| Batch size ($B$) | 32 |
| Epochs | Early stopping (`patience` $= 100$) |

### S2.3 DeepONet

The DeepONet architecture used as a baseline for comparison with the FNO is shown in Figure S3. DeepONet consists of a *branch network*, which takes the axial displacement field as input, and a *trunk network*, which takes the spatial coordinates (query points) at which the elastic modulus is to be predicted. Using the notation (batch size, channel, height, width), the branch network receives input of shape $(B, 1, 100, 100)$ and processes it through three convolution-SiLU (sigmoid linear unit) blocks, producing feature maps of shape $(B, 128, 13, 13)$. These feature maps are flattened and passed through fully connected layers to yield hidden features of shape $(B, 256)$.

The trunk network takes query points of shape $(B \times n_\mathrm{q}, 2)$, where $n_\mathrm{q}$ is the number of query points, and maps them through fully connected layers to produce hidden features of shape $(B \times n_\mathrm{q}, 256)$. In this study, all query points ($100 \times 100$ spatial coordinates) were used during training to maximize utilization of available information. The hidden features are then reshaped to $(B, n_\mathrm{q}, 256)$. Finally, the outputs of the branch and trunk networks are combined via an inner product to predict the elastic modulus at each $(x, y)$ location. The hyperparameter settings are summarized in Table S3.

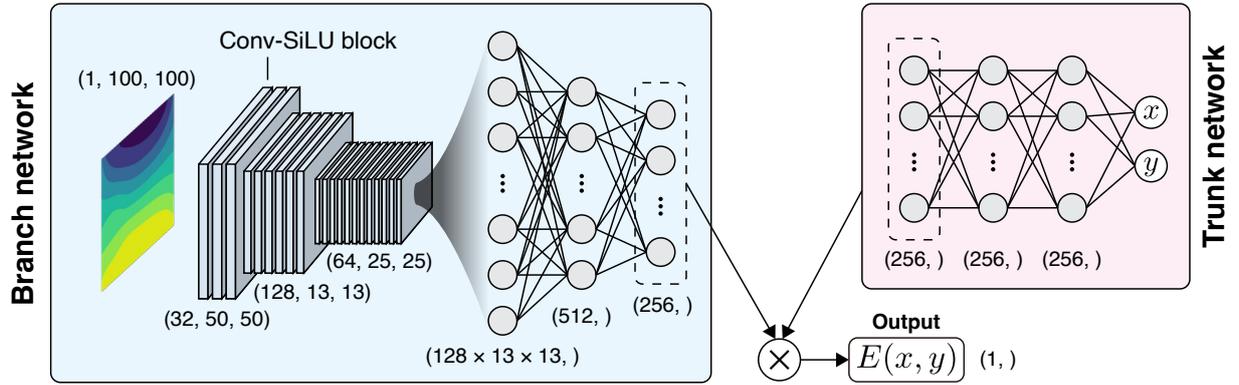

**Figure S3.** Architecture of the DeepONet used in this study.

**Table S3.** Hyperparameters of the DeepONet.

| Parameter | | Value |
|---|---|---|
| Branch network | No. of layers | 6 (3 convolution + 3 linear) |
| Trunk network | No. of layers | 3 linear |
| Training parameters | Normalizer | Gaussian normalizer |
| | Loss function | Eq. (5) (from the main text) |
| | Optimizer | Adam (`weight_decay` $= 1 \times 10^{-4}$) |
| | Learning rate | $1 \times 10^{-3}$ |
| | Learing rate scheduler | StepLR (`step_size` $= 100$, `gamma` $= 0.5$) |
| | Batch size ($B$) | 32 |
| | Epochs | Early stopping (`patience` $= 100$) |
| | Width | 4 |

## S3 Predictive performance of the FNO for various tumor types

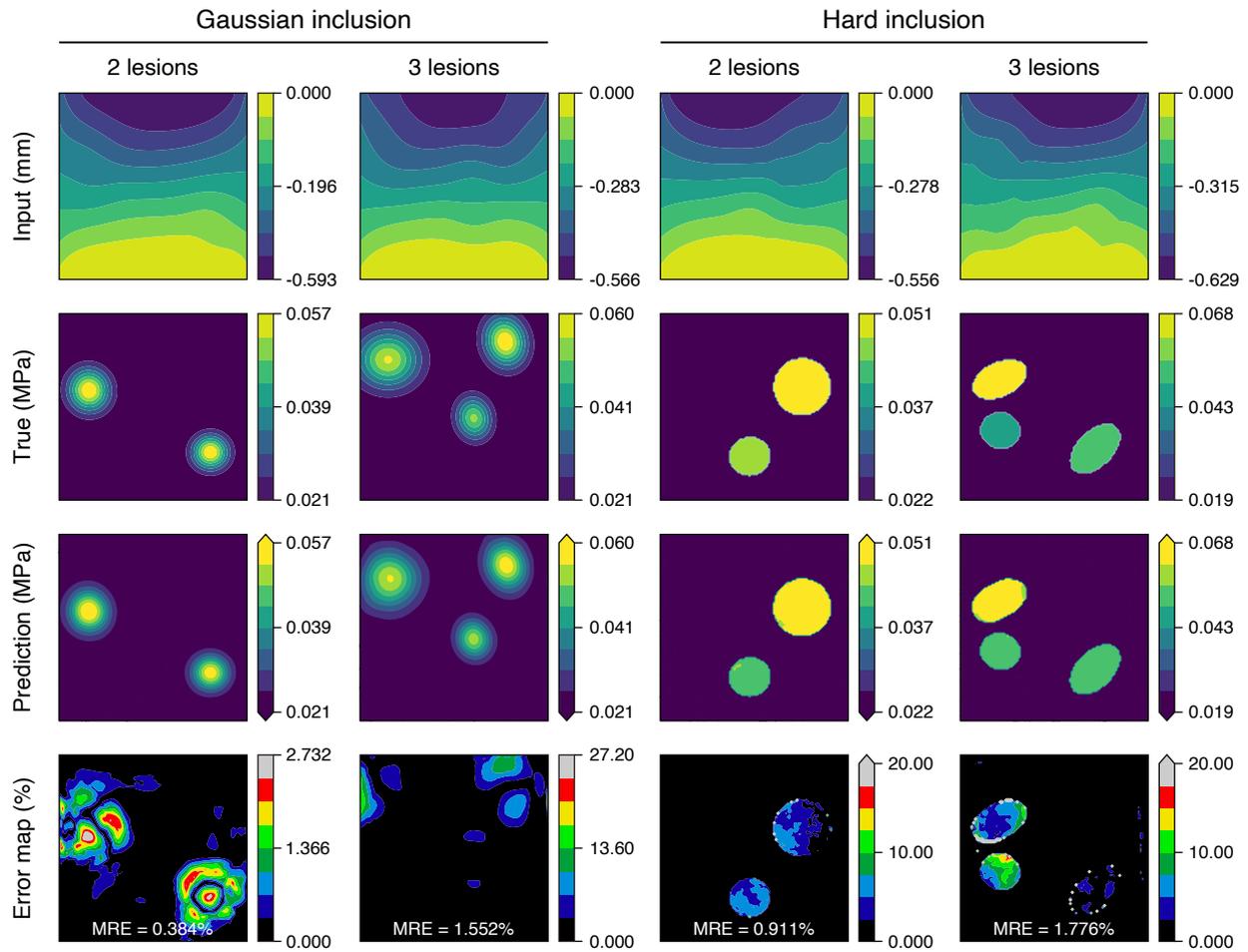

**Figure S4.** Predictions of the FNO for two and three Gaussian- or hard-inclusion-type lesions. For clearer visual comparison, the colorbar ranges of the ground truth and prediction are matched to the ground truth.

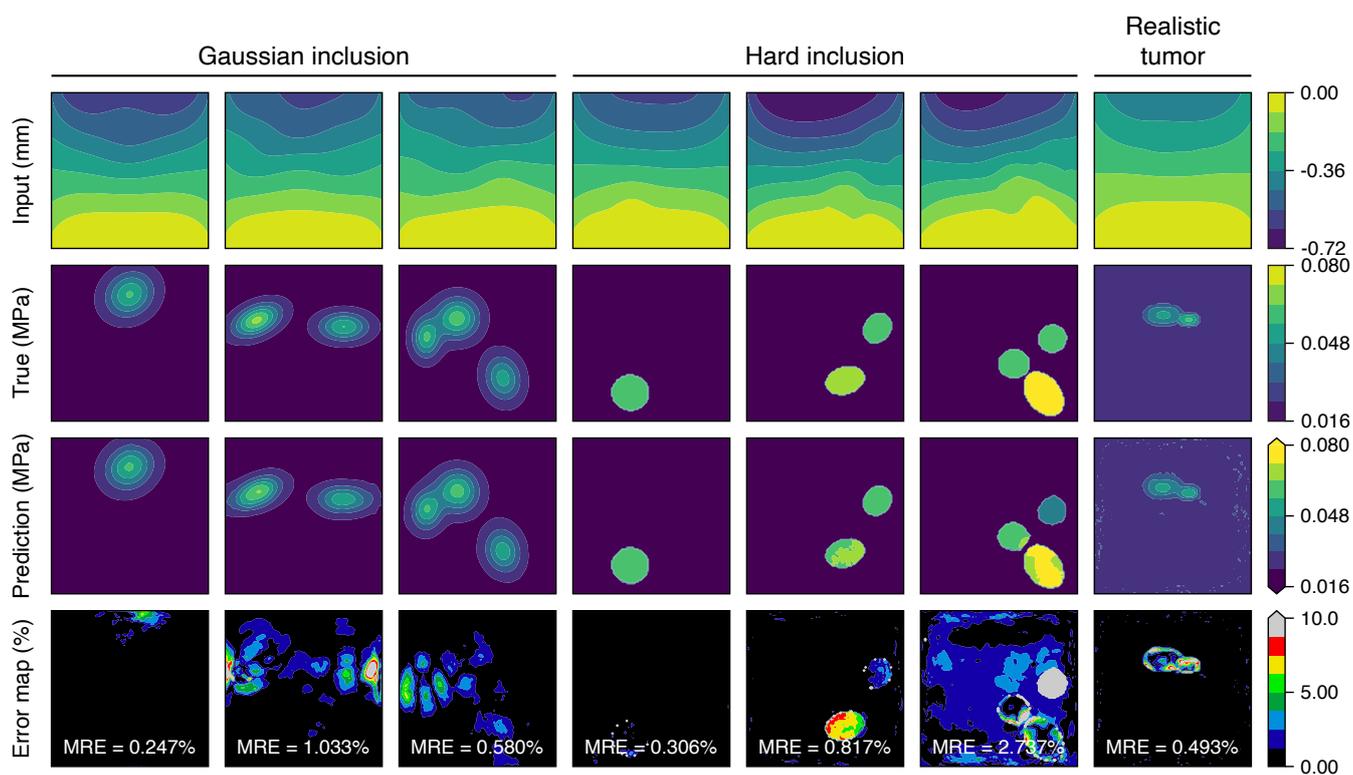

**Figure S5.** Predictions of the FNO for mixed dataset. For clearer visual comparison, the colorbar ranges of the ground truth and prediction are matched to the ground truth.